\documentclass[11pt,a4paper]{article}

\usepackage[T1]{fontenc}
\usepackage[utf8]{inputenc} 
\usepackage{graphicx}
\usepackage{amsmath,amssymb}
\usepackage{cite}
\usepackage[colorlinks=true,linkcolor=blue,citecolor=blue,urlcolor=blue]{hyperref}
\usepackage{orcidlink}
\usepackage{QCDEX2026}

\newcommand{\PI}{\texttt{HAPS-PiFF1.0}}
\newcommand{\KA}{\texttt{HAPS-KaFF1.0}}
\newcommand{\HA}{\texttt{HAPS-hFF1.0}}

\begin{document}

\noindent\includegraphics[width=0.4\textwidth]{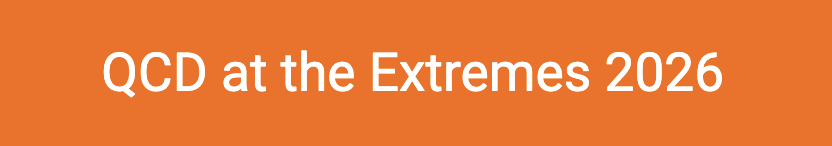}\par
\vspace{0.6em}
\begin{center}
  {\LARGE\bfseries Precision Fragmentation Functions from Modern COMPASS SIDIS Data\par}
  \vspace{0.8em}
  {\large Maryam Soleymaninia$^{1,*}$\,\orcidlink{0000-0002-2127-6933}\par}
  \vspace{0.35em}
  {\small On behalf of HAPS Collaboration\par}
  \vspace{0.35em}
  {\small $^{1}$School of Particles and Accelerators, Institute for Research in Fundamental Sciences (IPM), Tehran, Iran\par}
  \vspace{0.35em}
  {\small $^{*}$Corresponding author: \href{mailto:Maryam_Soleymaninia@ipm.ir}{Maryam\_Soleymaninia@ipm.ir}\par}
  \vspace{0.35em}
  {\small\itshape refereed by the conveners of QCDEX2026\par}
\end{center}
\vspace{0.5em}

\begin{abstract}
We present a comparative account of three global determinations of collinear fragmentation functions (FFs) for charged pions, charged kaons, and unidentified charged hadrons. The \PI{}, \KA{}, and \HA{} analyses combine single-inclusive $e^+e^-$ annihilation with charge-separated semi-inclusive deep-inelastic scattering (SIDIS), including the 2025 COMPASS proton-target measurements and the revised 2026 isoscalar-target results. Flexible neural-network parametrizations, Monte Carlo replicas, and NLO/NNLO perturbative setups are employed. The proton and isoscalar samples sharpen the flavor sensitivity of SIDIS, while gluon fragmentation remains comparatively weakly constrained. We discuss the effects of perturbative accuracy, the principal limitations, and prospects for future measurements.
\end{abstract}

\noindent\textbf{Keywords:} fragmentation functions; QCD global analysis; COMPASS; SIDIS; neural networks; NNLO

\section{Introduction}
Fragmentation functions $D_i^h(z,Q)$ describe the nonperturbative transition of a parton $i$ into a hadron $h$ with momentum fraction $z$. Their flavor decomposition is essential for collider and lepton--hadron phenomenology. Single-inclusive annihilation (SIA) provides precise quark-charge-weighted spectra, whereas SIDIS combines initial-state parton distributions (PDFs) with charge- and target-dependent information about the produced hadron~\cite{MAPFF}.

This contribution synthesizes the independent HAPS fits of $\pi^\pm$ and $K^\pm$~\cite{HAPS_piK} and unidentified $h^\pm$~\cite{HAPS_h}. The key experimental development is the COMPASS proton-target multiplicity measurement~\cite{COMPASS_proton} together with the 2026 revision of the earlier $^6$LiD-target results~\cite{COMPASS_iso}. The latter employs DJANGOH-based radiative corrections in place of the original TERAD treatment; it is a \emph{reanalysis} of existing isoscalar data rather than a new measurement. These complementary targets help distinguish up-, down-, and strange-quark fragmentation when fitted with SIA data.

\section{Formalism, data, and fitting strategy}
The SIDIS hadron multiplicity is the ratio of semi-inclusive to inclusive DIS cross sections,
\begin{equation}
 M_N^h(x,z,Q^2)=\frac{d\sigma_N^h/(dx\,dQ^2\,dz)}{d\sigma_N^{\rm DIS}/(dx\,dQ^2)},
 \label{eq:mult}
\end{equation}
with the numerator schematically $f_i^N\otimes C_{ij}^{\rm SIDIS}\otimes D_j^h$. The dependence on target $N$ and hadron charge gives SIDIS a flavor sensitivity unavailable to inclusive SIA. The pion and kaon extractions include SIA, HERMES~\cite{HERMES}, and COMPASS multiplicities; the $h^\pm$ extraction uses SIA and COMPASS. Inclusive $h^\pm$ FFs are fitted separately and cannot be approximated simply by the sum of pion and kaon FFs.

The complementary processes constrain different directions in flavor space. The $Z$-pole SIA measurements, including flavor-tagged samples, help determine heavy-quark contributions, while lower-energy SIA data test the scale dependence. In SIDIS, $h^+$ and $h^-$ multiplicities distinguish favored and unfavored fragmentation through their differing partonic weights. Combining proton with approximately isoscalar targets changes the relative up- and down-quark contributions, reducing degeneracies that are otherwise present in SIA alone. These advantages depend on the reliability of the PDF input and the treatment of correlated experimental uncertainties.

Timelike DGLAP evolution uses \textsc{APFEL++}, while NNPDF4.0 PDFs enter the SIDIS predictions~\cite{HAPS_piK,HAPS_h}. The analyses use $Q_0=5$~GeV, $\alpha_s(M_Z)=0.118$, $m_c=1.51$~GeV and $m_b=4.92$~GeV. At NNLO, we use the SIA coefficient functions and the SIDIS implementation adopted in \textsc{MontBlanc}~\cite{MontBlanc,Abele}; exact NNLO SIDIS calculations~\cite{Bonino} provide a benchmark for future systematic comparisons.

At the input scale, seven independent flavor combinations for each positively charged hadron are represented by a one-hidden-layer $[1,20,7]$ neural network, with
\begin{equation}
 zD_i^{h^+}(z,Q_0)=\big[N_i(z;\boldsymbol\theta)-N_i(1;\boldsymbol\theta)\big]^2,
 \label{eq:nn}
\end{equation}
which enforces positivity and the $z\!=\!1$ endpoint condition. The negatively charged FFs follow by charge conjugation. Monte Carlo data replicas are refitted to propagate experimental uncertainties; the central values and uncertainty bands correspond to the replica mean and standard deviation. The flavor bases and uncertainty correlations are documented in Refs.~\cite{HAPS_piK,HAPS_h}.

\begin{table}[!htbp]
\centering\small
\caption{Separate HAPS fits: number of retained SIA and SIDIS points, minimum SIDIS scale, and global fit quality. Values are taken from Refs.~\cite{HAPS_piK,HAPS_h}.}
\label{tab:fit}
\begin{tabular}{@{}lrrrrr@{}}
\hline
Final state & $N_{\rm SIA}$ & $N_{\rm SIDIS}$ & $Q_{\rm min}$ [GeV] & NLO & NNLO\\
\hline
$\pi^\pm$ & 377 & 918 & 1.5 & 0.87 & 0.87\\
$K^\pm$ & 339 & 795 & 1.7 & 1.13 & 0.89\\
$h^\pm$ & 370 & 886 & 1.4 & 0.943 & 0.873\\
\hline
\end{tabular}
\end{table}

The SIDIS lower-$Q$ cuts in Table~\ref{tab:fit} arise from fit-stability studies and should not be interpreted as eliminating all hadron-mass or power corrections. The COMPASS $h^\pm$ analysis retains $0.2<z<0.85$, while the identified-hadron analyses also impose dataset-dependent $z$ restrictions.

\section{Results and discussion}
\begin{figure}[!htbp]
 \centering
 \makebox[\linewidth][c]{\includegraphics[width=.5\linewidth]{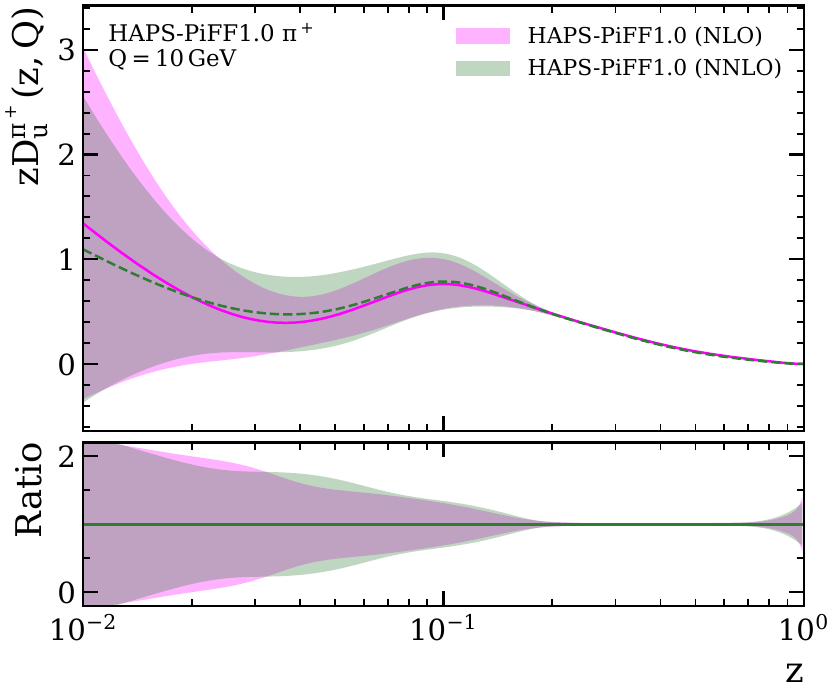}\hspace{.04\linewidth}\includegraphics[width=.53\linewidth]{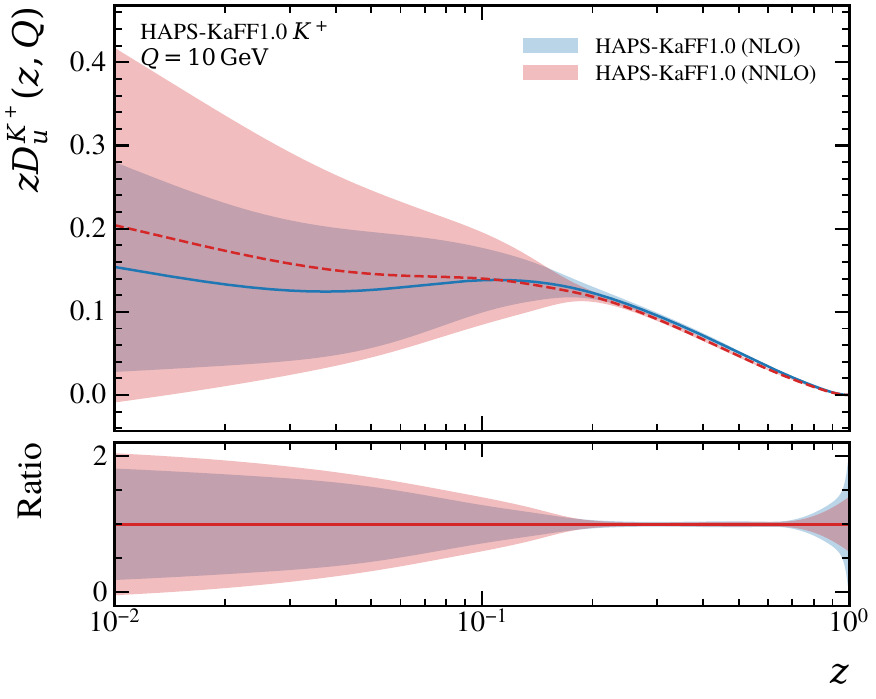}}\\[-1.5mm]
 \makebox[\linewidth][c]{\includegraphics[width=.53\linewidth]{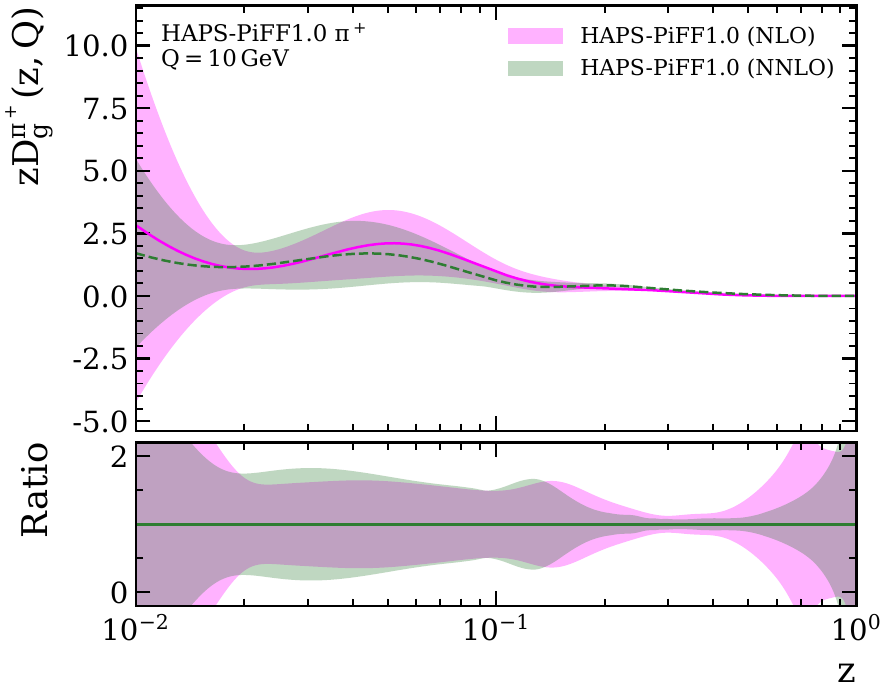}\hspace{.01\linewidth}\includegraphics[width=.55\linewidth]{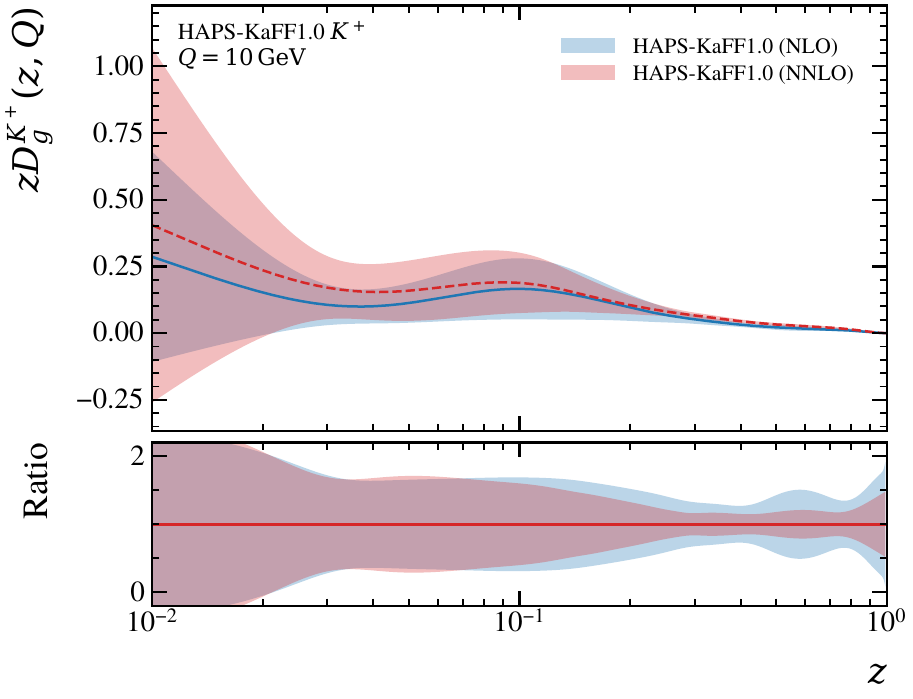}\hspace{.01\linewidth}}
 \caption{NLO and NNLO $u$-quark (top) and gluon (bottom) FFs for $\pi^+$ (left) and $K^+$ (right) at $Q=10$~GeV. Shaded regions show fitted replica uncertainties. Each lower subpanel displays the corresponding band relative to its \emph{own} central value, rather than the ratio of NNLO to NLO central predictions. Figures from the \PI{} and \KA{} analyses~\cite{HAPS_piK}.}
 \label{fig:piK}
\end{figure}

The three global fits describe the data well (Table~\ref{tab:fit}): the global
$\chi^2/N_{\rm dat}$ decreases from $1.13$ to $0.89$ for \KA{}, remains
$0.87$ (to two decimal places) for \PI{}, and decreases from $0.943$ to
$0.873$ for \HA{} when going from NLO to NNLO~\cite{HAPS_piK,HAPS_h}.
Table~\ref{tab:compass} compares the corresponding charge-combined COMPASS
subsets. For $K^\pm$, the proton (2025) subset improves from $1.41$ to
$1.17$, and the revised isoscalar (2026) subset from $1.11$ to $0.90$.
For $\pi^\pm$, the proton subset improves from $0.83$ to $0.74$, whereas
the isoscalar subset changes from $0.94$ to $1.01$. The combined $h^\pm$
proton subset remains approximately $0.83$ at both orders, while the
isoscalar subset improves from $1.04$ to $0.88$. Here $N_{\rm dat}$
counts the retained multiplicity bins, not the number of produced hadrons:
the 2025 (2026) samples contain $444$ ($442$) pion, $385$ ($394$)
kaon, and $444$ ($442$) unidentified-hadron bins, with positive and
negative charges combined. Species-dependent kinematic selections account
for the different sample sizes.Changes between perturbative orders also include refitting the
FFs, rather than corrections evaluated with fixed FFs.

\begin{table}[!htbp]
	\centering\small
	\caption{COMPASS subset $\chi^2/N_{\rm dat}$ and retained point counts,
		with positive and negative hadron charges combined within each target
		sample~\cite{HAPS_piK,HAPS_h}. The $h^\pm$ entries are
		$N_{\rm dat}$-weighted averages of the charge-separated values in
		Ref.~\cite{HAPS_h}, rounded to two decimal places.}
	\label{tab:compass}
	\begin{tabular}{@{}llrrr@{}}
		\hline
		Hadron & COMPASS target/data release & $N_{\rm dat}$ & NLO & NNLO\\
		\hline
		$\pi^\pm$ & Proton (2025)              & 444 & 0.83 & 0.74\\
		$\pi^\pm$ & Isoscalar, revised (2026)  & 442 & 0.94 & 1.01\\
		$K^\pm$   & Proton (2025)              & 385 & 1.41 & 1.17\\
		$K^\pm$   & Isoscalar, revised (2026)  & 394 & 1.11 & 0.90\\
		$h^\pm$   & Proton (2025)              & 444 & 0.83 & 0.83\\
		$h^\pm$   & Isoscalar, revised (2026)  & 442 & 1.04 & 0.88\\
		\hline
	\end{tabular}
\end{table}

Figure~\ref{fig:piK} shows the $u$-quark and gluon FFs at $Q=10$~GeV. For the quark channels, the NLO and NNLO central predictions largely overlap through the constrained $z$ region. By contrast, the gluon FFs are more uncertain, particularly at small $z$, because SIA and SIDIS probe the gluon mainly through higher-order hard coefficients and QCD evolution. The charge and target dependence of the COMPASS data improves access to favored and unfavored quark fragmentation; strange-to-kaon separation remains correlated with PDF and flavor-model choices. In particular, the kaon fit keeps the $\bar s\to K^+$ favored channel independent from $s\to K^+$, whereas the pion analysis allows the two favored charged-pion channels to differ. These choices are useful for testing flavor patterns but are not substitutes for genuinely independent measurements. The overlap of bands across perturbative orders is most meaningful in the kinematically constrained region; at endpoints, relative uncertainties become large and differences between fitted central values are not necessarily significant.

\begin{figure}[!htbp]
 \centering
 \makebox[\linewidth][c]{\includegraphics[width=.5\linewidth]{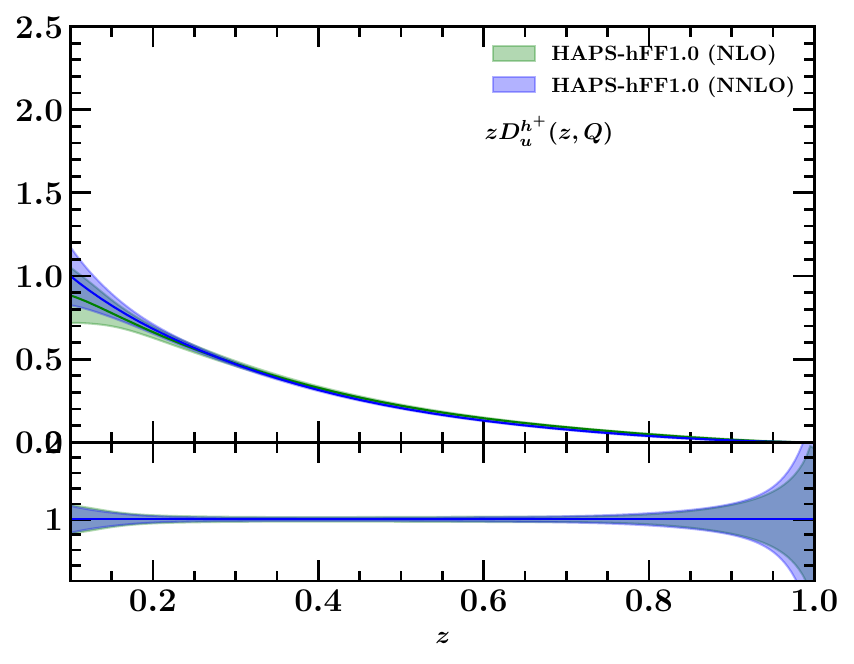}\hspace{.04\linewidth}\includegraphics[width=.5\linewidth]{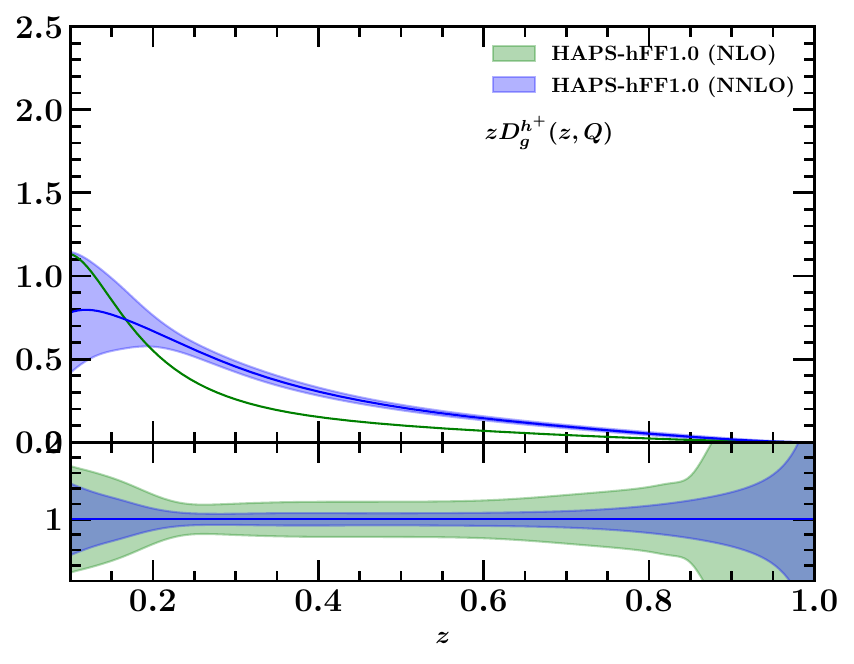}}
 \caption{\HA{} NLO (green) and NNLO (blue) determinations for $u\to h^+$ (left) and $g\to h^+$ (right), at $Q=5$~GeV. The lower panels contain self-normalized uncertainty bands, not NNLO/NLO ratios~\cite{HAPS_h}.}
 \label{fig:hFF}
\end{figure}

The unidentified-hadron results (Fig.~\ref{fig:hFF}) exhibit similar quark-sector perturbative stability and wider gluon uncertainties. The $h^\pm$ FFs characterize an inclusive charged-hadron final state, retaining contributions beyond pions and kaons; their separate extraction provides an additional consistency test of the joint SIA--SIDIS description.

The improved COMPASS radiative corrections put the revised isoscalar multiplicities on a more consistent methodological footing with the 2025 proton sample. Nevertheless, no matched refit replacing only the 2017 data by the 2026 revision is presented here: the difference cannot be interpreted as a separately quantified reduction in FF uncertainty. Replica bands capture the adopted experimental covariance and fit procedure, not the full set of possible perturbative, PDF, cut, or parametrization systematics. The $h^\pm$ analysis, for example, assigns $80\%$ of COMPASS systematic uncertainties to correlated contributions~\cite{HAPS_h}. The newer radiative treatment is especially relevant at low Bjorken $x$ and high hadron $z$, where the COMPASS addendum reports changes in correction factors reaching approximately $12\%$~\cite{COMPASS_iso}. This is a correction to the experimental extraction of multiplicities, not a universal $12\%$ shift in the fitted FFs. Because the analyzed data include relatively low $Q$, complementary variations of the scale cuts, parton distributions and hadron-mass/power corrections are necessary to determine how robust extrapolations remain. Gluon-sensitive hadron-collider final states would provide an additional handle largely orthogonal to the quark-flavor information supplied by SIDIS.

\section{Summary and outlook}
The \PI{}, \KA{}, and \HA{} determinations establish a consistent methodology across three different final states using modern charge-separated COMPASS SIDIS information. NNLO provides its clearest global improvement for charged kaons, whereas pion results remain globally stable and unidentified charged hadrons show a more moderate improvement. The quark FFs are reasonably stable in data-sensitive regions; gluon fragmentation continues to motivate additional measurements and theory tests.
The resulting
\texttt{HAPS-PiFF1.0}, \texttt{HAPS-KaFF1.0} and \texttt{HAPS-HFF1.0} replica sets are made
publicly available in the standard \textsc{LHAPDF} format
~\cite{HAPS-PiFF10,HAPS-KaFF10, HAPS-HFF10}.   
 Future studies should assess exact NNLO SIDIS implementations, and incorporate gluon-sensitive hadron-collider and Electron--Ion Collider observables.

\bibliographystyle{QCDEX2026}
\bibliography{QCDEX2026}

\end{document}